\documentclass[DIV=calc,paper=a4,fontsize=11pt,twocolumn]{scrartcl} 

\pdfoutput=1
\pdfmapfile{+txfonts.map}

\usepackage[english]{babel}
\usepackage[protrusion=true,expansion=true]{microtype}
\usepackage{amsmath,amsfonts,amsthm,amssymb}
\usepackage[final]{graphicx}
\usepackage{xcolor}
\usepackage[normal,small,hypcap,up,labelfont=bf,textfont=it]{caption}
\usepackage{subfig}
\usepackage{booktabs}
\usepackage{fix-cm}
\usepackage{dsfont}
\usepackage{bbm}
\usepackage{cite}
\usepackage[utf8]{inputenc}
\usepackage[perpage,symbol*]{footmisc}
\usepackage[varg]{txfonts}
\usepackage{fancyhdr}
\PassOptionsToPackage{hyphens}{url}\usepackage[pdfencoding=auto,psdextra]{hyperref}
\usepackage{bookmark}
\usepackage{verbatim}
\usepackage{fontenc}
\usepackage{cuted}
\usepackage{widetext}

\DeclareCaptionFont{mycolor}{\color[HTML]{000000}}
\usepackage{sectsty}													
\allsectionsfont{
\color[HTML]{31ADF3}\usefont{OT1}{phv}{b}{n}
}

\sectionfont{
\color[HTML]{31ADF3}\usefont{OT1}{phv}{b}{n}
}

\usepackage{fancyhdr}												
\usepackage{lettrine}
\newcommand{\initial}[1]{%
\lettrine[lines=3,lhang=0.3,nindent=0em]{
\color[HTML]{31ADF3}
{\textsf{#1}}}{}}

\usepackage{titling}															

\newcommand{\HorRule}{\color[HTML]{31ADF3}
\rule{\linewidth}{1pt}%
}

\pretitle{\vspace{-30pt} \begin{flushleft} \HorRule
\fontsize{34}{34} \usefont{OT1}{phv}{b}{n} \color[HTML]{31ADF3} \selectfont
}
\title{Can Decoherence Solve the Measurement Problem?}					
\posttitle{\par\end{flushleft}\vskip 0.5em}

\preauthor{\begin{flushleft}\large \lineskip 0.5em \usefont{OT1}{phv}{b}{sl} \color[HTML]{31ADF3}}
\author{Mani L. Bhaumik\\[8pt]}											
\postauthor{\footnotesize \usefont{OT1}{phv}{m}{sl} \color[HTML]{000000}
Department of Physics and Astronomy, University of California, Los Angeles, USA. E-mail: \href{mailto:bhaumik@physics.ucla.edu}{bhaumik@physics.ucla.edu}\\[10pt]		
\scriptsize\usefont{OT1}{phv}{m}{n} \color[HTML]{31ADF3}{\textbf{Editors: \emph{Zvi Bern} \& \emph{Danko Georgiev}} }\\[5pt]
\color[HTML]{000000}{Article history: Submitted on November 28, 2022;  Accepted on December 3, 2022; Published on December 4, 2022.}
\par\end{flushleft}\HorRule}

\date{}																				

\begin{document}
\maketitle
\thispagestyle{fancy} 			
\initial{T}\textbf{he quantum decoherence program has become more attractive in providing an acceptable solution for the long-standing quantum measurement problem. Decoherence by quantum entanglement happens very quickly to entangle the quantum system with the environment including the detector. But in the final stage of measurement, acquiring the unentangled pointer states poses some problems. Recent experimental observations of the effect of the ubiquitous quantum vacuum fluctuations in destroying quantum entanglement appears to provide a solution.\\ Quanta 2022; 11: 115--123.}

\begin{figure}[b!]
\rule{245 pt}{0.5 pt}\\[3pt]
\raisebox{-0.2\height}{\includegraphics[width=5mm]{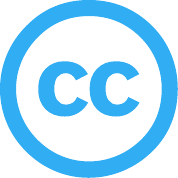}}\raisebox{-0.2\height}{\includegraphics[width=5mm]{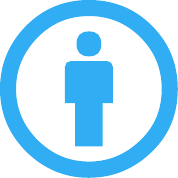}}
\footnotesize{This is an open access article distributed under the terms of the Creative Commons Attribution License \href{http://creativecommons.org/licenses/by/3.0/}{CC-BY-3.0}, which permits unrestricted use, distribution, and reproduction in any medium, provided the original author and source are credited.}
\end{figure}

\section{Introduction}

Shortly after the formulation of quantum mechanics nearly a century
ago, a rather strange aspect of the theory became apparent. It is
variously known as the \emph{quantum measurement problem}, \emph{wave function collapse} or \emph{wave function reduction}. This is because the non-relativistic quantum
mechanics appears to consist of two essentially different processes.
After a unitary evolution of the wave function of a microscopic quantum
state following the linear Schr\"{o}dinger equation, one must resort
to a sudden non-unitary stochastic collapse of the wave function to
obtain its classical measurement outcome.

The genesis of the problem started with the quantum pioneers headed
by Niels Bohr. They insisted that we only ever observe any physical
phenomena at the macroscopic level. We never directly deal with the
quantum objects of the microscopic realm and therefore need not worry
about them or their physical reality. Accordingly, they argued that
both the observer and the measurement apparatus must be kept outside
the system to which quantum mechanics is applied. This is known as
the Copenhagen interpretation, which simply pronounced the issue of
microscopic quantum states is out of bounds, stating that physicists
just had to accept a fundamental distinction between the quantum and
the classical domains. Without being disrespectful to the esteemed
founding fathers of quantum mechanics, we may inquire how can such
thoughts be ever consistent with a scientific outlook?

Nonetheless, this was epitomized by the mathematical mastermind John
Von Neumann in his classic axiomatic formulation of non-relativistic
quantum mechanics using the linear Hilbert vector space \cite{1b,1c}. Even after
many decades, his skillful formulation is still taught in almost all
advanced quantum mechanics classes despite the two obvious incongruities
arising from the Copenhagen interpretation. These comprise of the
essential \emph{ad hoc} role of consciousness and the postulated assumption
of an abrupt collapse of the wave function. Quantum mechanics itself
does not predict the collapse, which must be manually added to the
calculations. Einstein famously likened it to God playing dice to
decide what becomes ``real'' -- what we actually observe in our classical
world. However, despite the quantum pioneers' assertions, enormous
efforts by the physics community have been made leading to many alternate
postulates to explain away the irrational proposal of the inventors.
Significant progress has been made by these efforts but without leading
to any consensus, although some substantial fissures have ensued in
the original interpretation.

\section{The collapse postulate}

John von Neumann, way back in his formulation \cite{1b,1c}, postulated
his non-unitary ``process 1'', to emphasize the role of \emph{consciousness}
for the collapse of the wave function in the measurement process.
It was concluded by von Neumann and most of the physicists of the
time that there is no physical reason for the collapse in measurement
transition. Thus evolved the rather instinctive resort to the ``consciousness
of an external observer,'' which appears to be fading in time. Even
the stalwarts like Eugene Wigner fell for it \cite{Wigner1961} but
eventually repudiated it later \cite{Wigner1984}.

It is quite remarkable to note that so much effort by so many eminent
scientists in the early years of quantum mechanics was devoted to
the role of consciousness in quantum measurement. But this directly
contradicts the obvious fact that in the early years of the universe,
the conditions were not suitable for appearance of any manifest conscious
agents. Yet the universe developed to a mature state obeying quantum
rules long before the possibility of emergence of conscious beings.
This has been characterized by John Bell in jest
\begin{quote}
Was the wavefunction of the world waiting to jump for thousands of
millions of years until a single-celled living creature appeared?
Or did it have to wait a little longer, for some better qualified
system $\ldots$ with a PhD? \cite[p.~34]{2}
\end{quote}
To be fair, the conditions of the early universe were not known to
the pioneers of quantum physics. It would be reasonable to speculate
that very likely they would not have put such essential emphasis on
consciousness if they knew the early universe conspicuously ascertains
that consciousness is not essential for the workings of quantum rules.

The other enigmatic postulate that von Neumann institutionalized is
the collapse of the wave function initially alluded by Werner Heisenberg.
By then, it was well established that a quantum state in the macroscopic
domain is usually a superposition of two or more states. But in measurement
using classical devices, one observes only a single state and no superposition.
Von Neumann conjectured that in the measurement process, the quantum
states would collapse to one of the superposed states following his
improvised projection postulate. Again, in John Bell's words
\begin{quote}
If the theory is to apply to anything but highly idealised laboratory
operations, are we not obliged to admit that more or less `measurement-like'
processes are going on more or less all the time, more or less everywhere?
\cite[p.~34]{2}
\end{quote}
Gerhart L\"{u}ders rejected \cite{3} von Neumann's collapse postulate
(except for degenerate states). Its confirmation came in a recent
experiment performed by Pokorny \emph{et al}. \cite{4} who called
it the ideal measurement process. In this well-planned experiment,
the authors created a microscopic superposition of three quantum states.
They were able to measure just one of the superposed states without
collapsing the entire wave function also observing that the collapse
happened over time and not instantaneously. Serge Haroche and his group \cite{5}
also demonstrated that reduction of the wave function happens gradually.

An example from the cosmic history is worth examining in this regard.
The universe about 380 000 years after the big bang consisted primarily
of hydrogen ions (protons) and electrons, along with neutral helium
atoms. An electron would naturally be attracted to the proton, starting
to emit electromagnetic radiation due to its motion. But a much more
rapid process would take place when the electron, while aligned in
the direction of the proton, spontaneously emits a virtual photon
with an amount of energy that exactly matches the potential energy
of the electron in an orbital of the hydrogen atom. In this process,
the wave function of the electron can directly wind up as the wave
function of a specific orbital of the hydrogen atom without having
to undergo a typical collapse to any particular point. Such episodes
would reveal that the wave function does not necessarily always need
to go through a traditional collapse for detection.

But the mystery of the occurrence of the quantum to classical transition
continues to persist. Consequently, substantial attempts have been
made to find an acceptable solution by modifying the Schr\"{o}dinger
equation but without any success so far. Despite its outstanding success,
some experts like Vittorio Gorini, Andrzej Kossakowski, George Sudarshan
\cite{Gorini1976} and G\"{o}ran Lindblad \cite{Lindblad1976} have
attempted to modify the Schr\"{o}dinger equation to solve the measurement
problem. Steven Weinberg using the Lindblad equation pointed out \cite{6a,6b,6c}
using data from atomic clocks that any proposed modification would
need to produce an accuracy of at least one part in $10^{17}$ in
the difference between the energy states employed in the clock. The
accuracy of the atomic clocks continues to improve requiring possibly
even better improvement of the modification of the theory. So far,
such approaches do not seem to be fruitful.

\pagebreak
An attractive scheme generally known as the Ghirardi--Rimini--Weber
theory \cite{Ghirardi1986,7} has been studied extensively over the
last four decades by arbitrarily attaching a Gaussian function to
the Schr\"{o}dinger equation. The modification acts as a Markovian
process that has negligible effect during the unitary evolution but
becomes active afterwards during the final measurement when a very
large number of particles become available due to some unspecified
diffusion process. The efficacy of the Ghirardi--Rimini--Weber modified
Schr\"{o}dinger equation remains to be demonstrated. It may be prudent
to consider Steven Weinberg's contention
\begin{quote}
Unfortunately, these ideas about modification of quantum mechanics
are not only speculative but also vague, and we have no idea how big
we should expect the corrections to quantum mechanics to be. \cite[pp.~139--140]{6a}
\end{quote}
Roger Penrose has proposed \cite{8a,8b} a novel scheme of a gravitational
process to bring about the reduction of the wave function but without
any successful experimental demonstration yet. The most fruitful approach
now seems to be the one based on quantum decoherence. Furthermore,
the decoherence time being relatively short \cite{9}, also seem to
rule out the Ghirardi--Rimini--Weber modification and the gravitational
reduction proposal since both will take some time to be built up for
their effectiveness.

\section{Further progress}

It is rather amazing that not until about half a century after the
advent of quantum mechanics, Heinz-Dieter Zeh was the first to emphasize
that the microscopic quantum state wave function evolves unitarily
obeying the Schr\"{o}dinger equation in isolation from the environment
\cite{10}. However, for measurement, the wave function must be exposed
to the ambient atmosphere as well as to the plethora of quantum systems
of the measuring device. Under this open circumstance, the various
components of the superposed wave function become affected with the
elements of the environment as well as the measuring device.

This led to the initiation of a more systematic study of the effect
of the environment and the measuring apparatus on quantum system resulting
in the loss of quantum coherence, which is now known as \emph{decoherence}.
Use of density matrix was also initiated for decoherence by
Zeh in 1970s \cite{10,11}. Zeh continued his work on decoherence,
sometimes with Erich Joos, for decades \cite{12}. The next big step
forward came when the idea of quantum entanglement was conjoined with
decoherence for exploration of quantum measurement. It is fascinating
to appreciate how this historic conjunction came to be recognized.

The award of the 2022 Nobel Prize in physics has brought significant
attention to quantum entanglement. It is now well known that the essence
of quantum entanglement arose from the famous Einstein--Podolsky--Rosen
paper \cite{13} published way back in 1935. But it remained effectively
ignored by most physicists until John Bell's epochal article \cite{14}
on Bell's inequality proposed in 1964. Again it remained on the side
line for quite a while due to the lack of a suitable experimental
arrangement to verify Bell's proposal. Eventually a feasible experimental
arrangement was devised five years later by John Clauser, Michael
Horne, Abner Shimony and Richard Holt \cite{15}. The first experimental
verification of the Bell--Clauser--Horne--Shimony--Holt theory, now
popularly known as quantum entanglement, was carried out by Friedman
and Clauser in 1972 \cite{16} with later substantiation by Clauser
and Shimony \cite{17}. A lack of interest of the mainstream scientists
to the subject still continued perhaps because of possible loopholes
in the substantiation of Bell's theorem. The checkered history of the
development of this period included an underground journal to avoid
the apathy of \emph{Physical Review} editors to quantum entanglement.
This is well documented in a book amusingly entitled, \emph{How the
Hippies Saved Physics} and penned by David Kaiser \cite{18}.

Eventually, a better authentication of Bell's theorem came from Alain
Aspect and his group \cite{19,20}. Further experiments to provide
loophole free confirmation of Bell's theorem led to the acceptance
of quantum entanglement. An analysis of how does nature possibly accomplish
nonlocal action are presented by Bhaumik \cite{21}. The essential
role of entanglement in quantum decoherence was soon realized by K\"{u}bler
and Zeh \cite{22}. However, Zeh's emphasis of entanglement in further
studies of Everett's theory of quantum measurement apparently distracted
him from advancing the appropriate roles of entanglement in decoherence.
Nevertheless, he continued his work with others, Erich Joos being
one of them \cite{12}.

\section{Details of decoherence}

As soon as the closed quantum system is exposed to the environment
including the detector, the unitary Schr\"{o}dinger evolution in
a very short order generates quantum entanglement between the system
and the detector including the environment making it possible to combine
the system and the detector into a single bigger system. Since both
the system and the detector comprising atoms and molecules abide quantum
rules, one can build up a composite tensor product space using two
sets of orthonormal basis vectors of Hilbert spaces. The combined
system evolves in a unitary fashion. Outcome of the measurement of
the selected quantum system is determined by the quantum correlations
encoded in the globally entangled quantum states of the composite
system. Thus, the conspicuous feature of the decoherence program is
that the laws of quantum mechanics are not suspended during measurement,
contrary to the popular assumption of most of the pioneers of quantum
mechanics for a long time.

Since 1980, decoherence involving quantum entanglement has been extensively
studied by Wojciech Zurek with his group at the Los Alamos National
laboratory for almost four decades making a very substantial improvement
in our understanding of the process. In summary, Zurek's investigations
show that only the eigenstates or the pointer states survive in the
complex environmental decoherence process and the number of entanglement
states increases very substantially due to what Zurek calls quantum
Darwinism. Consequently, the plethora of entanglement states consisting
of the robust pointer states show up in the process of measurements.
Detailed mathematical analyses of the decoherence process have been
presented in a substantial number of publications by Zurek and others
\cite{24,25,26,27}. More recently, an excellent entire book on decoherence
has been presented by Maximillian Schlosshauer \cite{28}. A brief
synopsis of the essential results of all these investigations will
be presented next.

\section{Finding the expectation values}

Let us consider that the quantum system and the detector including
the environment are each represented by a finite dimensional Hilbert
space, $\mathcal{H}_{S}$ and $\mathcal{H}_{E}$, leading to a pure
composite state $\vert\psi_{SE}\rangle$ that can be represented by
a density matrix $\hat{\rho}_{SE}$ corresponding to the pure state
as
\begin{equation}
\hat{\rho}_{SE}=\vert\psi_{SE}\rangle\langle\psi_{SE}\vert.
\end{equation}
The expectation value $\langle\hat{A}\rangle$ of any observable $\hat{A}$
acting on $\mathcal{H}_{S}\otimes\mathcal{H}_{E}$ is
\begin{equation}
\langle\hat{A}\rangle=\textrm{Tr}\left(\hat{\rho}_{SE} \,\hat{A}\right),
\end{equation}
which is completely determined for the composite state.
Despite that the composite $\hat{\rho}_{SE}$ is pure,
in general, both the $\hat{\rho}_{S}$ and $\hat{\rho}_{E}$ individually
are ensemble of states. Each of their reduced density matrices contains
an incoherent mixture of $N$ quantum state vectors $\vert\psi_{n,i}\rangle$
\begin{equation}
\hat{\rho}_{i}=\sum_{n=1}^{N}p_{n,i}\vert\psi_{n,i}\rangle\langle\psi_{n,i}\vert
\end{equation}
where $i\in\{S,E\}$, $\vert\psi_{n,i}\rangle\langle\psi_{n,i}\vert$
are projection operators with probability $p_{n,i}$ and the sum of
the probabilities is normalized, $\sum_{n}p_{n,i}=1$. Thus, there can
be various ensembles of states with each one having its own probability
distribution that will produce the same density matrix. Therefore,
for a single copy of unknown state $\hat{\rho}_{SE}$, it is the case
that $\hat{\rho}_{i}$ are unknowable to any meaningful extent for
either of the components. However, if we are given multiple copies
of the same composite state $\hat{\rho}_{SE}$,
then $\hat{\rho}_{SE}$ and $\hat{\rho}_{i}$ can be reconstructed
using quantum state tomography \cite{tomo} and $\langle\hat{A}\rangle$
can be obtained as the average of measurement outcomes of~$\hat{A}$, where each measurement is performed on
a new copy of $\hat{\rho}_{SE}$. It is rather amusing to note that
we may know everything about the composite entangled pure state, while we may not
know anything specific for either one of the component mixed states.
Also, we may know exactly the expectation value of an observable
$\langle\hat{A}\rangle$, while we may not know what the measurement
outcome for each measurement run will be \cite{GG}.

In the situation when the system~$S$ and the environment~$E$ are
quantum correlated by entanglement, an observer having access only
to the system $S$ can compute the expectation values for any local
observable using only the system's reduced density matrix
\begin{equation}
\hat{\rho}_{S}=\textrm{Tr}_{E}\left(\hat{\rho}_{SE}\right)
\end{equation}
where the reduced density matrix $\hat{\rho}_{S}$ is obtained by
tracing out the degrees of freedom of the environment in the joint
system--environment density matrix $\hat{\rho}_{SE}$. The statistics
of all possible local measurements on the system $S$ is comprehensibly
encoded in the reduced density matrix. Thus, for any local observable
$\hat{A}_{S}\otimes\hat{I}_{E}$ that relates only to the Hilbert
space $\mathcal{H}_{S}$ of the quantum system $S$, the reduced density
matrix $\hat{\rho}_{S}$ will be sufficient to calculate the expectation
value of the observable
\begin{equation}
\langle\hat{A}_{S}\rangle=\langle\hat{A}_{S}\otimes\hat{I}_{E}\rangle=\textrm{Tr}\left(\hat{\rho}_{S}\hat{A}_{S}\right)
\end{equation}
Although the concept of the reduced density matrix was introduced
by Paul Dirac in 1930 \cite{Dirac1931}, oddly its significance does not appear to have
been appreciated for almost half a century until the advent of quantum
entanglement. An essential element of Zurek's milestone contributions
to the decoherence program turned out to be the utilization of entanglement
and consequently the reduced density matrix for dealing with expectation
values among others.

\section{The problem with decoherence}

Although Zurek and his colleagues have advanced the decoherence program
in leaps and bounds over the last four decades, there are still some
conspicuous complexities in resolving the measurement problem. To
begin with, although the trace rule provides a convenient way to obtain
the reduced density matrix and hence the expectation value of an observable,
the trace operation is a non-unitary process stroking a whiff of the
collapse theory.
More importantly, their work does not seem to provide a satisfactory
explanation of where does the probability in measurement come from.
Zurek's derivation has been criticized, among others, by Steven Weinberg.
In his classic textbook \emph{Lectures on Quantum Mechanics}, Weinberg
states
\begin{quote}
There seems to be a wide spread impression that decoherence solves
all obstacles to the class of interpretations of quantum mechanics,
which take seriously the dynamical assumptions of quantum mechanics
as applied to everything, including measurement. \cite[p.~88]{29}
\end{quote}
Weinberg goes on to characterize his objection by asserting that the derivation of Born's rule by Zurek is 
\begin{quote}
clearly circular,
because it relies on the formula for expectation values as matrix
elements of operators, which is itself derived from the Born rule.
\cite[p.~88]{29}
\end{quote}

Maximilian Schlosshauer has become a champion advocate of the application
of decoherence toward the resolution of the measurement problem among
others. In a paper on Zurek's derivation of the Born rule, he and Arthur Fine comment
\begin{quote}
Certainly Zurek's approach improves our understanding of the probabilistic
character of quantum theory over that sort of proposal by at least
one quantum leap. \cite{30}
\end{quote}
However, they also criticize Zurek's derivation of the Born rule of
circularity, stating
\begin{quote}
we cannot derive probabilities from a theory that does not already
contain some probabilistic concept; at some stage, we need to ``put
probabilities in to get probabilities out.'' \cite{30}
\end{quote}
In a recent paper \cite{31}, we have presented a plausible solution
that supplements decoherence with some basic aspects of the well-established
Quantum Field Theory of the Standard Model of Particle
Physics. Our argument relies on some characteristics of the universal
quantum fields that predetermine the values of the complex coefficients
involved in the inherent superposition of eigenstates before measurement.
This has been also briefly hinted by Leonard Susskind \cite{32} by
stating that the probability of a quantum state does not change during
unitary evolution, which is its attribute. Thus, one of the major
obstacles in using decoherence for quantum measurement could be considered
resolved.

The other significant problem is that although the reduced density
matrix gives a convenient way to find the expectation value of an
observable, unfortunately, decoherence does not provide the pointer
states separately. We only get those states still entangled with the
environment including the detector states and that is not what an
experimenter will measure. For that purpose, we need separable or
product states such as
\begin{equation}
\vert\psi_{S}\rangle\otimes\vert\psi_{E}\rangle
\end{equation}
We now present some plausible ways to accomplish this.

\section{Product states using quantum rules}

From the available facts so far, it appears fruitful to bring about
the innate existence of the ubiquitous \emph{vacuum quantum fluctuations}
for this objective. During the unitary evolution of some superposed
quantum states, no substantial effect of the fluctuations has been
observed other than their essential participation in \emph{spontaneous emission}.
The vital part played by the quantum fluctuations in facilitating
spontaneous emission, which is a unitary process according to quantum
electrodynamics, has been known from the early days of quantum mechanics.
It was conveyed in a recent article by the author \cite{31}, how
some additional properties of matter like the well-known Lamb shift,
anomalous $g$-factor, etc., would not exist without the ubiquitous
fluctuations of the electro-magnetic quantum fields. These quantum
fluctuations, essential for spontaneous emission, could very likely
separate the pointer states from the entanglement with the environment.

The quantum fluctuations are known to be represented by a Gaussian
function. The effect of the Gaussian quantum fluctuations has never
been witnessed to affect the unitary Schr\"{o}dinger evolution to
any appreciable degree. But its significant effect could be cumulatively
operative during the measurement process when a substantial number
of the entangled states have been produced. Thus, it seems reasonable
to explore if the quantum fluctuations could make the disentanglement
effective in aiding quantum measurement in somewhat of a manner envisioned
by the Ghirardi--Rimini--Weber proponents but without any modification
of the Schr\"{o}dinger equation as well as not requiring a very large
number of quantum states during the final measurement.

In our goal to understand the effect of quantum fluctuations to produce
disentanglement in recovering the product states, it seems prudent
to explore some of the relevant new activities being pursued by the quantum computation community.
After Peter Shor's publication
of his celebrated algorithm for quantum computing in 1994, extensive
studies have been carried out in both decoherence and disentanglement,
which is of critical importance to quantum technology for avoiding
loss of quantum coherence. Hence the activities on these topics have
exploded exponentially in the last two decades.
As a resource, quantum entanglement has now been measured, increased, decreased
or even distilled and teleported \cite{Nielsen2010}.

The necessity for investigating decoherence as well as disentanglement
for quantum technologies is opposite to our requirement in quantum
measurement but there could be a commonality. The quantum fluctuations,
so essential for spontaneous emission, could very well be involved
in terminating the entanglement with the detector states. Particularly,
it could be fruitful to pursue the surprising experimental observation
called ESD, which stands for \emph{early stage disentanglement} or
\emph{entanglement sudden death,} that has been observed by several
groups. In these experiments, astonishingly a very swift disappearance
of entanglement altogether has been reported \cite{33,34,35,36,37,38,39,40}.
Other works \cite{41} report entanglement breaking channels.

In the simplest experimental setup, two entangled atoms in their excited
states are placed one each in two widely separated cavities without
any direct interaction. When the two atoms reach their ground states
by spontaneous emission, surprisingly the entanglement suddenly disappears
completely and the two atoms in their ground state constitute product
states. Although not yet fully understood, the sudden disappearance
of the entanglement is an experimental fact that could possibly be
caused by a process like what occurs in the unitary quantum electrodynamical
depiction of spontaneous emission. If that turns out to be true, since
unitary process preserve the probability, the final reduced quantum
state would have the same probability all the way from superposition
to reduction. In contrast,
the von Neumann collapse postulate
assumes a non-unitary process following Born's rule.

The act of spontaneous emission appears to be a sudden non-unitary
jump, however, if one were to keep track of all the vacuum modes,
as per quantum electrodynamics the combined atom--vacuum system in
fact undergoes a unitary time evolution. Thus, there could be a plausible
chance that the ESD process might be unitary although the details
are not yet fully understood. Further studies are planned to explore
this propitious possibility.

Another feasible process resembling some aspect of the Ghirardi--Rimini--Weber
procedure appears promising. This is the experimentally observed disentanglement
caused by quantum fluctuations. Like a Markovian process, the quantum
fluctuations does not affect the normal Schr\"{o}dinger evolution
indicating a limitation of the strength of the relevant interaction.
It appears to take place only after enough states are made available
by quantum Darwinism, when the cumulative strength of interaction
would cause the disappearance of entanglement leading to the desired
separable states. However, since the same information about the pointer
observable is stored independently in many fragments of the environment,
suitable detectors can measure the observable in different fragments
even without any observer involved.

From the experimentally observed results of the consistent effect
of quantum fluctuations in diminishing quantum entanglement, this
approach appears to be assured for accomplishing the desired separable
states. Our goal is to capture the disentangled observables in the
detector. So we need to find out what could cause the disentanglement.
Several experiments clearly confirm that the vacuum quantum fluctuations
cause the disentanglement. Thus, the essential agent has been clearly
identified and we could leave at that. But the work would be more
complete if we can provide the rate and consequently the disentanglement
time that could be reasonably short.

We know that the quantum fluctuations can be represented by a Gaussian.
So we need to find the rates and value of the constants for the Gaussian
and then possibly making some calculations like in the Ghirardi--Rimini--Weber
model to predict the time taken for disentanglement. It is not essential
but would complete the program. However, because studies of the details
of the disentanglement process is still continuing vigorously and
many of the results does not clearly identify whether it was done
in a cavity where quantum electrodynamics can give more than a number
of rates say for spontaneous emission. Again, the most important part
is to experimentally identify the mechanism that causes disentanglement
and that we already have accomplished with reasonable confidence.
So the principal objective can be considered reasonably accomplished.

\section{Concluding remarks}

It is evident by now that the advent of quantum entanglement has led
to a quantum leap for a resolution of the enduring measurement problem
through the decoherence procedure. Additionally, the distinctive appeal
of this program revealing that the quantum rules are not suspended
during the measurement process is unique. Although many others have
contributed, the decades of concerted efforts by Zurek and his group
have advanced the progress of the decoherence program to a fairly
mature stage.

The primary deprecation of their advancement concerns, however, is
the lack of a satisfactory answer to the origin of probability and
the occurrence of separable product states in the measurement process.
A cogent perspective is presented here that appears to alleviate the
deficiency of achieving the expected observables and their probabilities
in measurement. Therefore, along with the prior article \cite{31}
by the author, a solution of the century old quantum measurement problem
could be on hand. Significantly, much of the process of the reduction
of the wave function or quantum to classical transition occur following
quantum rules in contrast to the visions of the pioneers of quantum
physics.

The universe is quantum at the core and so are we. About seven octillions
of electrons and a plethora of other elementary particles inhabit
our body. Our existence in the familiar classical world is made possible
by continuing transition from the quantum to classical domains additionally,
of course, with the irreversible metabolic processes. The quantum
origin of objects in the classical arena is patently supported by
the recent observation \cite{42} of a sliver of residual quantum
activity in a man size 40-kilogram mirror in the Laser Interferometer
Gravitational-Wave Observatory (LIGO). In fact, in a variety of experiments,
quantum effects have been observed from mesoscopic to macroscopic
entities clearly indicating a transition from quantum to classical
is the abiding rule when a quantum system is exposed to a huge number
of quantum particles \cite{43,44,45}.

Most significantly, deriving the wave function of a non-relativistic
quantum mechanics from the fundamental reality accessible to us so
far by the standard model of particle physics and utilized by the
author in a series of publications \cite{31,46,47,48}, illustrate
quantum mechanics could be considered weird no more. We must recognize
that there are two distinct parts of reality, the quantum and the
classical with their characteristic rules, but one transitioning to
the other. The perception of weirdness arise when we try to understand
our daily classical world through the lens of quantum rules. The quantum
theory could be as splendid a theory based on fundamental realty as
has been both the non-relativistic Newton's laws as well as Maxwell's
theory of electrodynamics.

\section*{Acknowledgments}

The author wishes to acknowledge helpful discussions with Zvi Bern
and Danko Georgiev.

\end{document}